\documentclass[a4paper]{jpconf}
\usepackage{graphicx}

\usepackage{amsmath,amssymb}

\def\ga{{\mbox {${\alpha}$}}}

\def\ge{{\mbox {${\epsilon}$}}}
\def\gm{{\mbox {${\gamma}$}}}
\def\gth{{\mbox {${\theta}$}}}
\def\vth{{\mbox {${\vartheta}$}}}
\def\lam{{\mbox {${\lambda}$}}}
\def\Gm{{\mbox {${\Gamma}$}}}
\def\Lam{{\mbox {${\Lambda}$}}}
\def\Sig{{\mbox {${\Sigma}$}}}
\def\pr{\prime}
\def\pd{\partial}
\def\half{\frac{1}{2}}
\def\upa{\uparrow}
\def\dwa{\downarrow}

\begin{document}
\title{Unified theory for external and internal attributes and symmetries of fundamental fermions}

\author{Ikuo S. Sogami}

\address{Maskawa Institute for Science and Culture, Kyoto Sangyo University, Kita-Ku, Kyoto 603-8555, JAPAN}

\ead{sogami@cc.kyoto-su.ac.jp}

\begin{abstract}
An unorthodox unified theory is developed to describe external and internal attributes and symmetries of fundamental fermions, {\it quarks and leptons}. Basic ingredients of the theory are an algebra which consists of all the triple-direct-products of Dirac $\gm$-matrices and a triple-spinor-field, called a {\it triplet field}, defined on the algebra. The algebra possesses three commutative subalgebras which describe, respectively, the external spacetime symmetry, the family structure and the internal color symmetry of quarks and leptons. The triplet field includes threefold (fourfold) repetitional modes of spin $\frac{1}{2}$ component fields with SU(3) (SU(4)) color symmetry. It is possible to qualify the Yukawa interaction and to make a new interpretation of its coupling constants naturally in an intrinsic mechanism of the triplet field formalism. The Dirac mass matrices with quasi-democratic structure are derived as an illustration.
\end{abstract}

\section{Introduction}
In the orthodox way of unification, such as the standard model and the grand unified theories, fundamental fermions belonging individually to the irreducible representations of the external Lorentz group are organized into definite multiplets of the internal symmetry group. Here, we propose an unorthodox scheme of unification where a family of fundamental fermions is described by a master spinor field obeying to a generalized Dirac equation with coefficients belonging to an algebra, called a {\it triplet algebra}, which consists of all the triple-direct-products of the Dirac matrices. In accordance with the Coleman-Mandula theorem~\cite{Coleman,Weinberg} insisting that {\lq\lq}spacetime and internal symmetries cannot be combined in any but a trivial way{\rq\rq}, we decompose the triplet algebra into mutually-commutative subalgebras. The subalgebra related with internal property of fundamental fermions, {\it quarks and leptons}, is made up of two subalgebras which can be related, respectively, to repetitional family structure and color symmetry. 
The master spinor field, named a {\it triplet field}, is proved to include threefold (fourfold) repetitional modes of spin $\frac{1}{2}$ component fields with SU(3) (SU(4)) color symmetry. 

We construct explicitly the triplet algebra and its subalgebra for the external spacetime symmetry in \S2 and define the triplet field forming a multiple representation of the Lorentz group in \S 3. Characteristic of the subalgebra for internal properties of fundamental fermions is closely examined in \S 4 and \S 5. First, chirality operators acting on the triplet field are decomposed further into the sums of the projection operators which can be interpreted to discriminate repetitional degrees of freedom for family structure. Second, we find the algebra for color symmetry in the set of the elements which are commutative both with the chirality operators and all elements of the algebra for external symmetry. 
In the present scheme, the triplet field has no degrees of freedom for the symmetry of electroweak interaction. In \S 6, we reformulate the standard model by introducing the set of chiral triplet fields for the left-handed doublet and the right-handed singlets of the Weinberg-Salam symmetry. It turns out possible to qualify the Yukawa interaction and make a new interpretation on the Yukawa coupling constants naturally in an intrinsic mechanism of the triplet field formalism. As an explicit illustration, the Dirac mass matrices with quasi-democratic structure are derived. Possible applications of the present scheme to the systems of fundamental fermions with different attributes and symmetries are discussed in \S 7.

\section{Triplet algebra and subalgebra for external Lorentz symmetry}
The Dirac's $\gm$-matrices $\gm_\mu\ (\mu = 0,\ 1,\ 2,\ 3)$ satisfying the Clifford anticommutation relations 
$\gm_\mu\gm_\nu + \gm_\nu\gm_\mu = 2\eta_{\mu\nu}1$ with
$(\, \eta_{\mu\nu}\, ) = {\rm diagonal}\, ( 1, -1, -1, -1 )$ are introduced solely as the mathematical quantities which carry no direct physical meaning. Hermite conjugate of $\gm_\mu$ is defined by $\gm_\mu^\dagger = \gm_0\gm_\mu\gm_0$. The $\gm$-matrices generate the 16 dimensional manifold, named a {\it Dirac algebra}, as
\begin{equation}
 A_{\gm} =\ \langle \gm_\mu \rangle =\ 
 \{1,\, \gm_\mu,\, \sigma_{\mu\nu},\, \gm_5\gm_\mu,\, \gm_5 \}
 \label{DiracAlg}
\end{equation}
where $\sigma_{\mu\nu} =\frac{i}{2}(\gm_\mu\gm_\nu-\gm_\nu\gm_\mu)$ and
$\gm_5 = i\gm_0\gm_1\gm_2\gm_3 =  \gm^5$.

Let us call the triple-direct-product of the bases $1,\, \gm_\mu,\, \sigma_{\mu\nu},\, \gm_5\gm_\mu$ and $\gm_5$ a {\it primitive triplet} and the triple-direct-product of arbitrary elements of $A_{\gm}$ a {\it triplet}. Then, we define the {\it triplet algebra} $A_T$ as the $16^3$ dimensional manifold spanned by all the linear combinations of triplets. In other words, the triplet algebra $A_T$ is generated in terms of the 12 primitive triplets $\gm_\mu\otimes 1\otimes 1,\ 
1\otimes\gm_\mu\otimes 1$ and $1\otimes 1\otimes\gm_\mu$ as follows:
\begin{equation}
   A_T = \langle \gm_\mu\otimes 1\otimes 1,\ 
   1\otimes\gm_\mu\otimes 1,\ 1\otimes 1\otimes\gm_\mu\rangle.
\end{equation}
The primitive triplets $A_1\otimes B_1\otimes C_1$ and $A_2\otimes B_2\otimes C_2$ obey the multiplication rule
\begin{equation}
  (A_1\otimes B_1\otimes C_1)(A_2\otimes B_2\otimes C_2)
  =A_1A_2\otimes B_1B_2\otimes C_1C_2
\end{equation}
and the triplets satisfy the distribution law
\begin{equation}
  \left({\sum_i} a_iA_i\right)\otimes
  \left({\sum_j} b_jB_j\right)\otimes
  \left({\sum_k} c_kC_k\right)
   = {\sum_{ijk}} a_ib_jc_k A_i\otimes B_j\otimes C_k
\end{equation}
where $A's$, $B's$ and $C's$ are the bases of $A_\gm$ in (\ref{DiracAlg}), and $a's$, $b's$ and $c's$ are complex numbers. The transpose (Hermite conjugate) of the primitive triplet is defined by the triple-direct-product of its transposed (Hermite conjugate) components of $A_\gm$ as
\begin{equation}
    \left(A\otimes B\otimes C\right)^{T}
      = A^{T}\otimes   B^{T}\otimes C^{T}, \quad
    \left(A\otimes B\otimes C\right)^\dagger
    =A^{\dagger}\otimes B^{\dagger}\otimes C^{\dagger}
\end{equation}
and the trace of the triplet is set to be
\begin{equation}
  {\rm Tr}\left(A\otimes B\otimes C\right)
           = {\rm Tr}\left(A\right) {\rm Tr}\left(B\right)
             {\rm Tr}\left(C\right).
\end{equation}

The triplet algebra $A_T$ is large enough to form a variety of realizations of the Lorentz group representation. To obtain such kind of multiple spinor representation that describes exclusively the spin $\frac{1}{2}$ states, we note that the primitive triplets
\begin{equation}
   \Gm_\mu = \gm_\mu\otimes\gm_\mu\otimes\gm_\mu\qquad
   (\mu = 0,\,1,\,2,\,3)
\end{equation}
satisfy the Clifford relations~\cite{Sogami}
\begin{equation}
   \Gm_\mu\Gm_\nu + \Gm_\nu\Gm_\mu = 2\eta_{\mu\nu}I
\end{equation}
where $I = 1\otimes 1\otimes 1$. Hermite conjugate of $\Gm_\mu$ is defined by
\begin{equation}
   \Gm_\mu^\dagger = \Gm_0\Gm_\mu\Gm_0  
\end{equation}
and the traces are calculated to be
\begin{equation}
  {\rm Tr}\,\Gm_\mu = 0,\ \ 
  {\rm Tr}\left( \Gm_\mu\Gm_\nu \right) = 64 \eta_{\mu\nu}.
\end{equation}
With these new elements, we are now able to construct a subalgebra of $A_T$ being isomorphic to the Dirac algebra $A_\gm$ as follows:
\begin{equation}
A_{\Gm} =\  \langle \Gm_\mu \rangle
=\ \{1,\, \Gm_\mu,\, \Sigma_{\mu\nu},\, \Gm_5\Gm_\mu,\, \Gm_5 \}
\end{equation}
where 
\begin{equation}
  \Sigma_{\mu\nu}=\frac{i}{2}(\Gm_\mu\Gm_\nu - \Gm_\nu\Gm_\mu)
 =-\sigma_{\mu\nu}\otimes\sigma_{\mu\nu}\otimes\sigma_{\mu\nu}
\end{equation}
and
\begin{equation}
  \Gm_5 = i\Gm_0\Gm_1\Gm_2\Gm_3 = \Gm^5
        = -\gm_5\otimes\gm_5\otimes\gm_5.
\end{equation}

The set of elements which are commutative with the subalgebra $A_\Gm$, {\it i.e.},
\begin{equation}
 C_\Gm =C(A_\Gm: A_T)=\{\,X \in A_T : [\, X,\ \Gm_\mu \,] =0\,\}
\end{equation}
forms a subalgebra of $A_T$, called a {\it centralizer} of $A_\Gm$ in $A_T$. The primitive triplet in $C_\Gm$ is proved to have even numbers of components $\gm_\mu$-matrices for all $\mu$~\cite{Sogami}. Therefore, the composition of $C_\Gm$ is determined to be
\begin{equation}
  C_\Gm = \langle\ 1\otimes \gm_\mu\otimes \gm_\mu,\  
   \gm_\mu\otimes 1\otimes\gm_\mu\ \rangle .
\end{equation}
The triplet algebra has a non-intersecting decomposition
$A_\Gm\cup C_\Gm =A_T$.

Note that all of the antisymmetric operators $\Sigma_{\mu\nu}$ form a subalgebra
\begin{equation}
A_\Sig = 
\left\{\raisebox{-0.3ex}{\normalsize$\Sig_{\mu\nu}$} \right\}\subset A_\Gm \subset A_T.
\end{equation}
We identify this subalgebra $A_\Sig$ with the {\it external algebra} which generates the Lorentz group in our spacetime. It is readily proved that the operators
\begin{equation}
   M_{\mu\nu} = \frac{1}{2}\Sigma_{\mu\nu}
\end{equation}
are subject to the commutation relations of the Lie algebra for the orthogonal group O(1,\,3) as
\begin{equation}
  [ M_{\kappa\lambda},\ M_{\mu\nu} ] = -i\eta_{\kappa\mu} M_{\lambda\nu}
        +i\eta_{\kappa\nu} M_{\lambda\mu}
        -i\eta_{\lambda\nu} M_{\kappa\mu} +i\eta_{\lambda\mu} M_{\kappa\nu} 
\end{equation}
and $M_{\mu\nu}$ and $\Gm_\lam$ satisfy the relations
\begin{equation}
  [ M_{\mu\nu},\ \Gm_\lambda ] = i\eta_{\lambda\nu} \Gm_{\mu}
                               -i\eta_{\lambda\mu} \Gm_{\nu} .
\end{equation}
Accordingly, it is possible to postulate that the operators $M_{\mu\nu}$ generate the Lorentz transformations and that the subscripts of operators consisting of $\Gm_\mu$ refer directly to the superscripts of the spacetime coordinates $\{\,x^\mu\,\}$ of the 4 dimensional world where we exist as observers. Therefore, it is natural to define the raising and lowering operations of index of $\Gm_\mu$ by the metric $\eta_{\mu\nu}$ as follows:
\begin{equation}
   \Gm^\mu = \eta^{\mu\nu}\Gm_\nu,\quad
   \Gm_\mu = \eta_{\mu\nu}\Gm^\nu .
\end{equation}

As is evident from the construction so far, the component $\gm$-matrices carry no direct physical meaning and regarded as playing only the role of {\it alphabets}. It is the elements of the triplet algebra $A_T$ that play the role of {\it codons} carrying the direct physical meanings.

\section{Multiple spinor representation of the Lorentz group}
At this stage, we are allowed to introduce the triplet field, $\Psi(x)$, on the spacetime point $x^\mu$, which spans a $4^3$ dimensional vector space for multiple spinor representations of the Lorentz group. The triplet field is proved to possess $4^2$ component fields with spin $\half$ carrying different internal attributes. For the triplet field $\Psi(x)$ and its adjoint field
\begin{equation}
  \overline{\Psi}(x) = \Psi^\dagger(x)\Gm_0,
\end{equation}
the scalar product is defined as
\begin{equation}
  \overline{\Psi}(x)\Psi(x)
  = \sum_{abc}\overline{\Psi}_{abc}(x)\Psi_{abc}(x).
\end{equation}
Under the proper Lorentz transformation
$x^{\prime \mu} = \Omega^\mu{}_\nu x^\nu$ where
$\Omega_{\lambda\mu}\Omega^\lambda{}_\nu = \eta_{\mu\nu}$
and ${\rm det}\,\Omega = 1$, the triplet field and its adjoint are presumed to transform as
\begin{equation}
  \Psi^\prime(x^\prime) = S(\Omega)\Psi(x),\ \ 
  \overline{\Psi}^\prime(x^\prime) = \overline{\Psi}(x)S^{-1}(\Omega)
\end{equation}
where the transformation matrix $S(\Omega)$ belongs to the external algebra $A_\Sig$. For the bilinear form $\overline{\Psi}(x)\Gm_\mu\Psi(x)$ to behave as the vector under the Lorentz transformation, $S(\Omega)$ must satisfy the conditions
\begin{equation}
  S^{-1}(\Omega) \Gm_{\mu} S(\Omega) = \Omega_\mu{}^{\nu}\Gm_\nu,
  \quad S^{-1}(\Omega) = \Gm_0 S^\dagger(\Omega) \Gm_0
\end{equation}
which determine its explicit form to be
\begin{equation}
   S(\Omega) =
   \exp\left(-\frac{i}{2}M_{\mu\nu}\omega^{\mu\nu}\right)
\end{equation}
for the spacetime rotations with angles $\omega^{\mu\nu}$ in the $\mu$-$\nu$ planes. For the discrete spacetime transformations such as the space inversion, the time reversal and the charge conjugation, the present scheme retains exactly the same structure as the ordinary Dirac theory.

The internal attributes of fundamental fermions should be fixed independently of the inertial frame of reference in which observations are made. This means that the generators specifying their internal attributes must be commutative with the elements of the external algebra $A_\Sig$ for the Lorentz transformations. Accordingly, we are led naturally to postulate that the algebra for internal symmetry should be related to the centralizer of $A_\Sig$ defined by
\begin{equation}
   C_\Sig = C(A_\Sig: A_T) = \{ X \in A_T : [\, X,\ \Sigma_{\mu\nu} \,] = 0 \}.
\end{equation}
The primitive triplet of the centralizer $C_\Sig$ is proved to consist of either an even number or an odd number of component $\gm_\mu$-matrix for all $\mu$~\cite{Sogami}. Consequently, the composition of the centralizer $C_\Sig$ is determined as follows:
\begin{equation}
   C_\Sig = \langle\ 1\otimes \gm_\mu\otimes \gm_\mu,\  
   \gm_\mu\otimes 1\otimes\gm_\mu,\ \gm_5\otimes\gm_5\otimes\gm_5\ \rangle .
\end{equation}

It is now recognized that the triplet algebra has another non-intersecting decomposition $A_\Sig\cup C_\Sig =A_T$. We have to examine detailed structures of $C_\Sig$, which is identified with {\it internal algebra}, and to inquire what sorts of internal attributes of the fundamental fermions can be inscribed on the triplet field $\Psi(x)$ forming the multiple spinor representation of the Lorentz group.
The subalgebras of the triplet algebra $A_T$ defined so far, $A_\Gm$, $A_\Sig$, $C_\Gm$ and $C_\Sig$, obey the inclusion relations
\begin{equation}
   A_\Sig \ \subset \ A_\Gm \  \subset \  A_T : \ \ 
   C_\Gm \ \subset \ C_\Sig \  \subset \  A_T.
\end{equation}
Connectedly, the element $\Gm_5$ belongs to the subalgebras
$A_\Gm$ and $C_\Sig$ as follows:
\begin{equation}
   \Gm_5 \notin A_\Sig, \  \ \Gm_5 \in A_\Gm :\ \ 
   \Gm_5 \notin C_\Gm, \ \ \Gm_5 \in C_\Sig.
   \label{whereisGm5}
\end{equation}

\section{Subdivision of chirality operators and repetitional modes of family structure}
Equation (\ref{whereisGm5}) shows that the chirality operators
\begin{equation}
   L = \frac{1}{2}(I-\Gm_5), \ \ R = \frac{1}{2}(I+\Gm_5)
\end{equation}
belong to both of the subalgebras $A_\Gm$ and  $C_\Sig$.
It is straightforward to verify that these operators have the following subdivisions in the internal algebra $C_\Sig$ as
\begin{equation}
  L = \ell\otimes r\otimes r + r\otimes\ell\otimes r
    + r\otimes r\otimes\ell + \ell\otimes\ell\otimes\ell
  \label{Loperator}
\end{equation}
and
\begin{equation}
  R = r\otimes \ell\otimes\ell + \ell\otimes r\otimes\ell
    + \ell\otimes\ell\otimes r + r\otimes r\otimes r 
  \label{Roperator}
\end{equation}
where
\begin{equation}
   \ell = \frac{1}{2}(1-\gm_5),\ \ r = \frac{1}{2}(1+\gm_5).
\end{equation}
Note that this subdivision cannot be realized in the algebra $A_\Gm$.

To give independent physical roles and meanings to those subdivided terms
in (\ref{Loperator}) and (\ref{Roperator}), let us introduce the following operators in $C_\Sig$ as
\begin{equation}
 \left\{
  \begin{array}{ll}
  \Pi_{1L} = \ell\otimes r\otimes r,\ &
  \Pi_{1R} = r\otimes \ell\otimes\ell, \\
 \noalign{\vskip 0.2cm}
  \Pi_{2L} = r\otimes\ell\otimes r,\ &
  \Pi_{2R} = \ell\otimes r\otimes\ell, \\
 \noalign{\vskip 0.2cm}
  \Pi_{3L} = r\otimes r\otimes\ell,\ &
  \Pi_{3R} = \ell\otimes\ell\otimes r,  \\
 \noalign{\vskip 0.2cm}
  \Pi_{4L} = \ell\otimes\ell\otimes\ell,\ &
  \Pi_{4R} = r\otimes r\otimes r, \\
  \end{array}
 \right.
\end{equation}
which are chiral projection operators obeying the relations
\begin{equation}
   \Pi_{ih}\Pi_{jh^\prime} = \delta_{ij}\delta_{hh^\prime}\Pi_{ih},\ \ \sum_{ih}\Pi_{ih}=I 
\end{equation}
for $i,\,j = 1,\,2,\,3,\,4$ and $h,\,h^\pr = L,\,R$. Next, we define the non-chiral projection operators
\begin{equation}
   \Pi_i = \Pi_{iL} + \Pi_{iR} 
\end{equation}
satisfying the relations
\begin{equation}
   \Pi_i\Pi_j = \delta_{ij}\Pi_i, \ \ \sum_{i}\Pi_{i}=I.
\end{equation}
All of these operators form the subalgebra $\{\Pi_{ih}\}=\langle\ \gm_5\otimes 1\otimes 1,\ 1\otimes\gm_5\otimes 1,\ 1\otimes 1\otimes\gm_5\ \rangle$ of the internal algebra $C_\Sig$.

Here, we make interpretation that the projected modes of triplet field given by
\begin{equation}
  \Psi_{iL}(x) = L\Pi_i\Psi(x) = \Pi_{iL}\Psi(x),\quad 
  \Psi_{iR}(x) = R\Pi_i\Psi(x) = \Pi_{iR}\Psi(x) 
\end{equation}
and
\begin{equation}
  \Psi_i(x) = \Pi_i\Psi(x) = \Psi_{iL}(x) + \Psi_{iR}(x)
\end{equation}
represent the repetitional family structures of quarks and leptons. To apply this scheme to the three generations of quarks and leptons observed in the low energy regime, we impose the constraint condition, a posteriori,
\begin{equation}
  \Pi_4\Psi(x) = 0 
  \label{constraint}
\end{equation}
on the triplet field to suppress the fourth mode. Henceforth, we describe this three generation model in the present paper. Note, however, that it is possible also to describe the model including the fourth mode whose existence could be confirmed in sufficiently high energy regime.

\section{Ordinary and extended color symmetries}
In the previous section, one aspect of the internal algebra $C_\Sig$ was clarified by examining its subalgebra $\{\Pi_{ih}\}$. Consequently, it is necessary and reasonable to investigate the structure of the centralizer $C(\,\{\Pi_{ih}\} : C_\Sig\,)$ of $\{\Pi_{ih}\}$ in $C_\Sig$. For its purpose, let us notice that the Dirac algebra $A_{\gm}$ has the three
elements
\begin{equation}
  \rho_1 = i\gm_2\gm_3,\ \rho_2 = i\gm_3\gm_1,
  \ \rho_3 = i\gm_1\gm_2
  \label{rho}
\end{equation}
which satisfy the multiplication rules of the Pauli algebra, {\it i.e.},
$\rho_a\rho_b = \delta_{ab}I + i\ge_{abc}\rho_c$.
Taking the triple-direct-products of these elements, we are able to construct the bases for a 16 dimensional submanifold included exclusively in the centralizer $C(\,\{\Pi_{ih}\} : C_\Sig\,)$ as follows:
\begin{equation}
\left\{\ 
\begin{array}{l}
    \lam_1 = \half\left(\rho_1\otimes\rho_1\otimes 1
          + \rho_2\otimes\rho_2\otimes 1\right), \qquad\ 
    \lam_2 = \half\left(\rho_1\otimes\rho_2\otimes\rho_3
          - \rho_2\otimes\rho_1\otimes\rho_3\right),\\
          \noalign{\vskip 0.3cm}
    \lam_3 = \half\left(1\otimes\rho_3\otimes\rho_3
          - \rho_3\otimes 1\otimes\rho_3\right),  \\
          \noalign{\vskip 0.3cm}
    \lam_4 = \half\left(\rho_1\otimes 1\otimes\rho_1
          + \rho_2\otimes 1\otimes \rho_2\right), \qquad\ 
    \lam_5 = \half\left(\rho_1\otimes\rho_3\otimes\rho_2
          - \rho_2\otimes\rho_3\otimes\rho_1\right),\\
          \noalign{\vskip 0.3cm}
    \lam_6 = \half\left(1\otimes\rho_1\otimes\rho_1
          + 1\otimes\rho_2\otimes\rho_2\right), \qquad\ 
    \lam_7 = \half\left(\rho_3\otimes\rho_1\otimes\rho_2
          - \rho_3\otimes\rho_2\otimes\rho_1\right),\\
          \noalign{\vskip 0.3cm}
    \lam_8 = \frac{1}{2\sqrt{3}}\left(1\otimes\rho_3\otimes\rho_3
          + \rho_3\otimes 1\otimes\rho_3 
          -2\rho_3\otimes\rho_3\otimes 1\right), \\
          \noalign{\vskip 0.3cm}
    \lam_9 = \half\left(1\otimes \rho_1\otimes\rho_1
           - 1\otimes\rho_2\otimes\rho_2\right), \qquad\ 
    \lam_{10} = -\half\left(\rho_3\otimes\rho_1\otimes\rho_2
          + \rho_3\otimes\rho_2\otimes\rho_1\right),\\
          \noalign{\vskip 0.3cm}
    \lam_{11} = \half\left(\rho_1\otimes 1\otimes\rho_1
          - \rho_2\otimes 1\otimes \rho_2\right), \qquad
    \lam_{12} = -\half\left(\rho_1\otimes\rho_3\otimes\rho_2
          + \rho_2\otimes\rho_3\otimes\rho_1\right),\\
          \noalign{\vskip 0.3cm}
    \lam_{13} = \half\left(\rho_1\otimes\rho_1\otimes 1
          - \rho_2\otimes\rho_2\otimes 1\right), \qquad
    \lam_{14} = \half\left(\rho_1\otimes\rho_2\otimes\rho_3
          + \rho_2\otimes\rho_1\otimes\rho_3\right),\\
          \noalign{\vskip 0.3cm}
    \lam_{15} =-\frac{1}{\sqrt{6}}\left(1\otimes\rho_3\otimes\rho_3
          + \rho_3\otimes 1\otimes\rho_3 
          + \rho_3\otimes\rho_3\otimes 1 \right). \\
\end{array}
\right.
\label{gencolorsym}
\end{equation}
The 15 operators $\lam_j$ ($j = 1,\ 2,\ \cdots ,15$) satisfy the commutation relations
\begin{equation}
        [\lam_j,\ \lam_k] = 2f_{jkl}\lam_l
        \label{comsu4}
\end{equation}
and the anticommutation relations
\begin{equation}         
        \{\lam_j,\ \lam_k\} = \delta_{jk}I + 2d_{jkl}\lam_l
\end{equation}
of the Lie algebra A$_3$, where $f_{jkl}$ and $d_{jkl}$ are the
symmetric and antisymmetric constants characterizing the algebra. The operators $\lam_j$ are self-adjoint and have the traces
\begin{equation}
     {\rm Tr}\lam_j = 0,\quad {\rm Tr}\lam_j\lam_k = 32\delta_{jk}.
\end{equation}
We postulate, here, that the Lie group generated by the algebra
\begin{equation}
     A_{c4} = \{I, \lam_1,\ \lam_2,\ \cdots ,\lam_{15} \}
\end{equation}
describes an extended color SU$_c$(4) symmetry. Evidently, this is one realization of the Pati-Salam symmetry in which the leptons are interpreted as the 4-th {\it lilac} color mode.

The projection operators to the quark and lepton sectors are constructed, respectively, by
\begin{equation}
   \Lam^{(q)} = \frac{1}{4}\left(3I - \otimes\rho_3\otimes\rho_3
   - \rho_3\otimes 1\otimes\rho_3 - \rho_3\otimes\rho_3\otimes 1\right)
\end{equation}
and
\begin{equation}
\Lam^{(\ell)} = \frac{1}{4}\left(I + \otimes\rho_3\otimes\rho_3
   + \rho_3\otimes 1\otimes\rho_3 + \rho_3\otimes\rho_3\otimes 1\right)
   \equiv \Lam_{\ell}
\end{equation}
which are subject to the relations
\begin{equation}
       \Lam^{(a)}\Lam^{(b)} = \delta_{ab}\Lam^{(a)}
\end{equation}
for $a, b = q, \ell$. Accordingly, the operator for {\it baryon number minus lepton number} takes the form
\begin{equation}
    B-L = \frac{1}{3}\Lam^{(q)} - \Lam^{(\ell)}
        = - \frac{1}{3}(1\otimes\rho_3\otimes\rho_3
          + \rho_3\otimes 1\otimes\rho_3 
          + \rho_3\otimes\rho_3\otimes 1)
        = \sqrt{\frac{2}{3}}\lam_{15}.
    \label{BL15}
\end{equation}
It is straightforward to verify that the operator $\Lam^{(q)}$ has the subdivision as
\begin{equation}
      \Lam^{(q)} = \Lam_r + \Lam_y + \Lam_g
\end{equation}
where
\begin{equation}
 \left\{\ 
  \begin{array}{l}
   \Lam_r = \frac{1}{4}\left(I + \otimes\rho_3\otimes\rho_3
   - \rho_3\otimes 1\otimes\rho_3 - \rho_3\otimes\rho_3\otimes 1
   \right), \\
   \noalign{\vskip 0.3cm}
   \Lam_y = \frac{1}{4}\left(I - \otimes\rho_3\otimes\rho_3
   + \rho_3\otimes 1\otimes\rho_3 - \rho_3\otimes\rho_3\otimes 1
   \right), \\
   \noalign{\vskip 0.3cm}
   \Lam_g = \frac{1}{4}\left(I - \otimes\rho_3\otimes\rho_3
   - \rho_3\otimes 1\otimes\rho_3 + \rho_3\otimes\rho_3\otimes 1
   \right)\\
     \noalign{\vskip 0.1cm}
  \end{array}
  \right.
\end{equation}
are the projection operators into tricolor quark modes.
The operators $\Lam_a$ satisfy the relations
\begin{equation}
   \Lam_a\Lam_b = \delta_{ab}\Lam_a, \ \ \sum_{a}\Lam_{a}=I
\end{equation}
for $a,\,b = r,\,y,\,g,\,\ell$.

The extended color symmetry SU$_c$(4) is presumed to be broken spontaneously at an appropriate energy scale down to the symmetry of semi-simple group SU$_c$(3)$\otimes$U$_{B-L}$(1). To see how SU$_c$(3) is embedded in SU$_c$(4), it is necessary to note the following property of the constant $d_{jkl}$ of SU(4) group as $d_{jkl} = \frac{1}{\sqrt{6}}\delta_{k15}$ 
for $j\leq 8$, and the relation between $\lam_{15}$ and $\Lam^{(q)}$ in (\ref{BL15}). For the eight operators $\lam_j$\ $(j=1,\ 2,\ \cdots, 8)$, we obtain the commutation relations in (\ref{comsu4}) with the structure constants $f_{jkl}$ of the SU(3) group and anticommutation relations
\begin{equation}
       \{\lam_j,\ \lam_k\} = \frac{4}{3}\delta_{jk}\Lam^{(q)}
        + 2d_{jkl}\lam_l .
\end{equation}
Note that the projection operator $\Lam^{(q)}$ appears in the right hand side of this relation. Without it, this relation runs into contradiction, when applied to the operator $\Lam^{(\ell)}$ from the right. The identities
\begin{equation}
        \Lam^{(a)}\lam_j = \delta_{aq}\lam_j
\end{equation}
which hold for $a=q, \ell$ and $j=1, 2, \cdots, 8$ imply that $\lam_j (\leq 8)$
are simultaneous eigenvectors of $\Lam^{(q)}$ and $\Lam^{(\ell)}$ with
respective eigenvalues 1 and 0, and that the operators $\lam_j (\leq 8)$ annihilate
the leptonic mode. These results prove that the 9 dimensional submanifold of the
extended color algebra $A_{c4}$ defined by
\begin{equation}
       A_{c3} = \{I, \lam_1, \lam_2, \cdots, \lam_8 \}
\end{equation}
generates the color SU$_c$(3) group. The U$_{B-L}$(1) group is generated by the operator $B-L$ given in (\ref{BL15}).

From the triplet spinor field $\Psi(x)$, the quark and lepton fields are projected out by
\begin{equation}
  \Psi^{(a)}(x) = \Lam^{(a)}\Psi(x),
\end{equation}
respectively, for $a = q,\,\ell$, and four color modes of SU$_c$(4) symmetry are given by
\begin{equation}
  \Psi_a(x) = \Lam_a\Psi(x)
\end{equation}
for $a = r,\,y,\,g,\,\ell$.

\section{Standard model in the triplet field formalism}
To separate the component fields of internal modes of the triplet field, it is convenient to introduce the braket symbols for the projection operators $\Lam_a$ and $\Pi_{ih}$
by
\begin{equation}
    \Lam_a = \mid a\rangle\langle a\mid,\ \ 
    \Pi_{ih} = \mid i\,h\rangle\langle i\,h\mid .
\end{equation}
Then, the decomposition of the bilinear scalar and vector forms of the triplet fields can be achieved as follows:
\begin{equation}
  \bar{\Psi}(x)\Psi(x)
  =\sum_{a}\sum_{ih}\bar{\Psi}(x)\Lam_a\Pi_{ih}\Psi(x)
  =\sum_{a}\sum_{ih}\bar{\Psi}_{ai\bar{h}}(x)\Psi_{aih}(x)
  \label{scalarbillinear}
\end{equation}
and
\begin{equation}
  \bar{\Psi}(x)\Gm_\mu\Psi(x)
  = \sum_{a}\sum_{ih}\bar{\Psi}(x)\Gm_\mu\Lam_a\Pi_{ih}\Psi(x)
  = \sum_{a}\sum_{ih}\bar{\Psi}_{aih}(x)\Gm_\mu\Psi_{aih}(x)
  \label{vecterbillinear}
\end{equation}
where $\bar{h}$ implies that $\bar{L}=R$ and $\bar{R}=L$. The chiral component fields $\Psi_{aih}(x)$ and $\bar{\Psi}_{aih}(x)$ are defined by
\begin{equation}
  \Lam_a\Pi_{ih}\Psi(x) = |aih\rangle\langle aih|\Psi(x)\rangle
                        = |aih\rangle\Psi_{aih}(x)
\end{equation}
and
\begin{equation} 
 \bar{\Psi}(x)\Lam_a\Pi_{ih} = 
 \langle\bar{\Psi}(x)|aih\rangle\langle aih|
 = \bar{\Psi}_{ai\bar{h}}(x)\langle aih|.
\end{equation}
We have extracted maximally $4^2$ component spinor fields with spin $\half$, $\Psi_{ai}=\sum_h\Psi_{aih}$, from the original triplet spinor field $\Psi(x)$.

In the present formalism, the triplet field has no degrees of freedom for the Weinberg-Salam symmetry SU$_L(2)\otimes$U$_Y$(1).
Therefore, to describe the actual system of quarks and leptons
in this algebraic scheme, it is required to prepare a set of triplet fields to make up for the deficit degrees of freedom for electroweak interaction. 

Here we incorporate the Weinberg-Salam symmetry into our formalism by introducing the set of chiral triplet fields so that the left-handed triplets constitute the doublet
\begin{equation}
  \Psi_L(x) =
     \left(  
       \begin{array}{ccc}
          \Psi^{(\upa)}(x)\\
          \noalign{\vskip 0.2cm}
          \Psi^{(\dwa)}(x)\\
       \end{array}
     \right)_L
  \label{chiraldoublet}
\end{equation}
and the right-handed triplets form the singlets
\begin{equation}
          \Psi_R^{(\upa)}(x),\quad \Psi_R^{(\dwa)}(x)
  \label{chiralsinglets}         
\end{equation}
of the electroweak isospin, where $(\upa)$ and $(\dwa)$ signify the up and down states. The quark and lepton parts of the electroweak doublet $\Psi_L(x)$ have the component modes expressed schematically as
\begin{equation}
  \Psi_L^{(q)}(x) = \Lam^{(q)}\Psi_L(x) =
     \left(  
       \begin{array}{ccc}
          \Lam^{(q)}\Lam_a\Pi_{iL}\Psi^{(\upa)}\\
          \noalign{\vskip 0.2cm}
          \Lam^{(q)}\Lam_a\Pi_{iL}\Psi^{(\dwa)}\\
       \end{array}
     \right)_L =
     \left(  
       \begin{array}{ccc}
        u_a\ & c_a\ & t_a\\
        \noalign{\vskip 0.2cm}
        d_a\ & s_a\ & b_a
       \end{array}
     \right)_L
\end{equation}
and
\begin{equation}
  \Psi_L^{(\ell)}(x) = \Lam^{(\ell)}\Psi_L(x) =
    \left(  
     \begin{array}{ccc}
       \Lam^{(\ell)}\Pi_{iL}\Psi^{(\upa)}\\
       \noalign{\vskip 0.2cm}
       \Lam^{(\ell)}\Pi_{iL}\Psi^{(\dwa)}\\
     \end{array}
    \right)_L
  = \left(  
     \begin{array}{ccc}
      \nu_e\ & \nu_\mu\ & \nu_\mu\\
      \noalign{\vskip 0.2cm}
      e\ & \mu\ & \tau
     \end{array}
    \right)_L .
\end{equation}
Similarly, the electroweak singlets $\Psi_R^{(\upa)}(x)$ and $\Psi_R^{(\dwa)}(x)$ possess the quark and lepton parts with contents of the component modes as follows:
\begin{equation}
\ \Psi_R^{(u)}(x) = \Lam^{(q)}\Psi_R^{(\upa)}(x)
 = \left(\Lam^{(q)}\Lam_a\Pi_{iR}\Psi_R^{(\upa)}\right)
 = \left(  
    \begin{array}{ccc}
     u_a\ & c_a\ & t_a\\
    \end{array}
   \right)_R ,
\end{equation}
\begin{equation}
\ \Psi_R^{(d)}(x) = \Lam^{(q)}\Psi_R^{(\dwa)}(x)
 = \left(\Lam^{(q)}\Lam_a\Pi_{iR}\Psi_R^{(\dwa)}\right)
 = \left(  
    \begin{array}{ccc}
     d_a\ & s_a\ & b_a\\
   \end{array}
   \right)_R ,
\end{equation}
\begin{equation}
  \Psi_R^{(\nu)}(x) = \Lam^{(\ell)}\Psi_R^{(\upa)}(x)
 = \left(\Lam^{(\ell)}\Pi_{iR}\Psi_R^{(\upa)}\right)
 = \left(  
    \begin{array}{ccc}
     \nu_e\ & \nu_\mu\ & \nu_\tau\\
    \end{array}
   \right)_R ,\ \ 
\end{equation}
and
\begin{equation}
\Psi_R^{(e)}(x) = \Lam^{(\ell)}\Psi_R^{(\dwa)}(x)
 = \left(\Lam^{(\ell)}\Pi_{iR}\Psi_R^{(\dwa)}\right)
 = \left(  
    \begin{array}{ccc}
     \ e\,\ \ & \mu\,\ \ & \tau\\
    \end{array}
   \right)_R .\ \ 
\end{equation}

The chiral triplets in (\ref{chiraldoublet}) and (\ref{chiralsinglets}) enable us to express the Lagrangian density of the kinetic part including the gauge interactions of quarks and leptons in the generic form
\begin{equation}
  {\cal L}_{kg} = \bar{\Psi}_L(x)\Gm^\mu{\cal D}_\mu\Psi_L(x)
  + \bar{\Psi}_R^{(\upa)}(x)\Gm^\mu{\cal D}_\mu\Psi_R^{(\upa)}(x)
  + \bar{\Psi}_R^{(\dwa)}(x)\Gm^\mu{\cal D}_\mu\Psi_R^{(\dwa)}(x)
\end{equation}
with the covariant derivatives ${\cal D}_\mu$. In the low energy regime where the standard model can be applied, the covariant derivatives act on the triplet fields as follows:
\begin{equation}
{\cal D}_\mu\Psi_L(x)=
\left(\pd_\mu - gA_\mu^{a}(x)\half\tau_a
              - g^{\pr}B_\mu(x)\half(B-L)
              - g_cA_{c\mu}^{a}(x)\half\lam_a \right)\Psi_L(x),
\end{equation}
\begin{equation}
{\cal D}_\mu\Psi_R^{(\upa)}(x)=
\left(\pd_\mu - g^{\pr}B_\mu(x)\half(B-L+1)
       - g_cA_{c\mu}^{a}(x)\half\lam_a \right)\Psi_R^{(\upa)}(x)
\end{equation}
and
\begin{equation}
{\cal D}_\mu\Psi_R^{(\dwa)}(x)=
\left(\pd_\mu - g^{\pr}B_\mu(x)\half(B-L-1)
      - g_cA_{c\mu}^{a}(x)\half\lam_a \right)\Psi_R^{(\dwa)}(x)
\end{equation}
where $A_\mu^{a}(x)$  and $B_\mu(x)$ are the gauge fields which interact, respectively, to the electroweak isospin $\half\tau_a$ of SU$_L(2)$ and the hypercharge of U$_Y$(1), and $A_{c\mu}^{a}(x)$ is the gauge field of color symmetry. This density ${\cal L}_{kg}$ is readily proved to reproduce the gauge interactions of the standard model with right-handed neutrino species. 

In the present scheme, the chiral triplet fields include the repetitional degrees of freedom for family structure. This feature enables us to formulate the Yukawa interactions in an intrinsic mechanism without bringing in so many unknown parameters from outside. In terms of chiral triplet fields, the Lagrangian density of the Yukawa interactions can be expressed to be
\begin{equation}
  \begin{array}{ll}
   {\cal L}_Y &= g_u\bar{\Psi}_L^{(q)}(x){\cal Y}_u
            \tilde{\phi}(x) \Psi_R^{(u)}(x)
            + g_d\bar{\Psi}_L^{(q)}(x){\cal Y}_d
            \phi(x)\Psi_R^{(d)}(x)\\
             \noalign{\vskip 0.3cm}
             &+\ g_\nu\bar{\Psi}_L^{(\ell)}(x){\cal Y}_\nu
            \tilde{\phi}(x)\Psi_R^{(\nu)}(x)
              + g_e\bar{\Psi}_L^{(\ell)}(x){\cal Y}_e
            \phi(x)\Psi_R^{(e)}(x) + {\rm h.c.}
  \end{array}
\label{Yukawaint}
\end{equation}
where
\begin{equation}
     \phi(x) =
      \left(
       \begin{array}{c}
        \phi^+ \\
        \noalign{\vskip 0.2cm}
        \phi^0 \\
       \end{array}
      \right),
      \quad
      \tilde{\phi}(x) =
      \left(
       \begin{array}{c}
        \phi^{0\ast} \\
        \noalign{\vskip 0.2cm}
        -\phi^- \\
       \end{array}
      \right)      
\end{equation}
are the Higgs field and its conjugate of the standard model, and
the kernel operator ${\cal Y}_a\ (a=u,\,d,\,\nu,\,e)$ works to induce mixings among repetitional modes. To be commutative with the generators of color symmetry, the operator must be composed of triple-direct-products of the Dirac matrices $\gm_0$ and $\gm_5$. To frame its generic construction, we impose the conditions that it is self-adjoint and it brings forth the quasi-democratic mass matrices in the low energy regime. As a natural candidate for ${\cal Y}_a$, we adopt
\begin{equation}
 \begin{array}{ll}
  {\cal Y}_a &= \half\left\{
  I + y_{a1} \left(1\otimes\gm_{0}e^{i\gth_{a2}\gm_5}\otimes
  \gm_{0} e^{i\gth_{a3}\gm_5}\right)\right. \\
     \noalign{\vskip 0.3cm}
 &\left.
 +\ y_{a2} \left(\gm_{0}e^{i\gth_{a1}\gm_5}\otimes
  1\otimes\gm_{0}e^{i\gth_{a3}\gm_5}\right)
 + y_{a3} \left(\gm_{0}e^{i\gth_{a1}\gm_5}\otimes\gm_{0}
  e^{i\gth_{a2}\gm_5}\otimes 1\right) \right\}\\
 \end{array}
 \label{kernel}
\end{equation}
with real parameters $y_{ai}$ and $\gth_{ai}\ (a=u,\,d,\,\nu,\,e;\ i=1,\,2,\,3)$. 

The quark part of the Lagrangian density in (\ref{Yukawaint}) is readily decomposed into the Yukawa interactions among component quark fields as
\begin{equation}
  \sum_{a}g_a\bar{\Psi}_L^{(q)}{\cal Y}_a\phi\Psi_R^{(a)} =
  \sum_{a}g_a\sum_{ij}\bar{\Psi}_L^{(q)}\Pi_{iR}
  {\cal Y}_a\Pi_{jR}\phi \Psi_R^{(a)} =
  \sum_{aij}\bar{\Psi}_{iL}^{(q)}
  \left[g_a\langle iR\mid{\cal Y}_a\mid jR\rangle\right]\phi
  \Psi_{jR}^{(a)}.
\end{equation}
This result and the similar decomposition of the lepton part lead us to the interpretation that the quantities
\begin{equation}
      g_a\langle iR|{\cal Y}_a|jR\rangle
\end{equation}
are the Yukawa coupling constants of the $a$ sector $(a=u,\,d,\,\nu,\,e)$.

In the low energy regime where the Higgs field takes the vacuum expectation value $\langle\phi\rangle$ and, consequently, the Weinberg-Salam symmetry is broken, the Yukawa interaction with the kernel in (\ref{kernel}) results in the Dirac mass matrices
\begin{equation}\hspace{-1.4cm}
 {\cal M}_a =
 \left( g_a\langle\phi\rangle\left(\langle iR\mid{\cal Y}_a\mid jR\rangle \right) \right) 
=m_a\left(
 \begin{array}{ccc}
  1  & \ga_{a3}e^{ i\vth_{a3}} &  \ga_{a2}e^{-i\vth_{a2}} \\
  \ga_{a3}e^{-i\vth_{a3}} & 1  &  \ga_{a1}e^{ i\vth_{a1}} \\
  \ga_{a2}e^{ i\vth_{a2}} & \ga_{a1}e^{-i\vth_{a1}} & 1   \\
 \end{array}
\right)
\end{equation}
where $m_a=\half g_a\langle\phi\rangle$, $\ga_{ai}=y_{ai}|\langle \ell|\gm_0|r\rangle|^2$ and $\vth_{a1}=\gth_{a2}-\gth_{a3}$ (cyclic). This type of the Hermitian quasi-democratic matrices are recognized to explain the hierarchical structures of mass spectra and the weak mixing matrix for the quark sector~\cite{Branco,SogamiNTS}.

\section{Discussion}
In this way, we have formulated an algebraic theory for unified description of external and internal symmetries of quarks and leptons. The basic ingredients of the theory are the {\it triplet algebra} consisting of all the triple-direct-products of the Dirac matrices and the {\it triplet field} forming the multiple spinor representation of the Lorentz group. The triplet field has the $4^3$ component fields which afford the freedom for the repetitional family structure as well as the external spacetime symmetry and the internal color symmetry.

By postulating the triplet fields of the electroweak doublet and singlets to exist, the standard model with three generations is successfully reproduced so that the Yukawa interactions are formulated, without bringing in many unknown coupling constants from outside, and the Dirac mass matrices with quasi-democratic structure are naturally derived.

The present formalism is flexible enough to be applicable to various field theories suitable for descriptions for different energy scales. Removing of the constraint in (\ref{constraint}) enables us formulate the four generation theory of quarks and leptons and introduction of the right-handed electroweak doublet in place of the singlets in (\ref{chiralsinglets}) allows us to obtain the gauge theory of left-right symmetry. Further, we can bring in the gauge symmetry for repetitional family modes. Note that the three Dirac matrices $\rho_1^{\pr}=\gm_0$, $\rho_2^{\pr}=i\gm_0\gm_5$ and $\rho_3^{\pr}=\gm_5$ generate the Pauli algebra. Taking the triple-direct-products of these elements just as in (\ref{gencolorsym}), we can construct the 16 dimensional submanifold which generates the SU(4) group for family symmetry. Therefore, it is possible to formulate the gauge field theory with the symmetry SU$_c(4)\otimes$SU$_L(2)\otimes$SU$_R(2)\otimes$SU$_{\rm family}$(4) in the triplet field formalism.

\ack
This work was supported by the JSPS Grant-in-Aid for Scientific Research (S) No.\,22224003.

\end{document}